\documentclass[]{spie}  

\usepackage[]{graphicx}

\title{Inference of stochastic nonlinear oscillators
with applications  to physiological problems}

\author{Vadim N. Smelyanskiy\supit{a}, Dmitry G. Luchinsky\supit{b}
 \skiplinehalf
\supit{a} NASA Ames Research Center, Mail Stop 269-2, Moffett
Field, CA 94035, USA; \\
\supit{b} Department of Physics, Lancaster University, Lancaster
LA1 4YB, UK. }

\authorinfo{Further author information: (Send correspondence to V.N.S.)\\V.N.S.:
E-mail: vadim@email.arc.nasa.gov}

\begin{document}
  \maketitle

\begin{abstract}
A new method of inferencing of coupled stochastic nonlinear
oscillators is described. The technique does not require extensive
global optimization, provides optimal compensation for
noise-induced errors and is robust in a broad range of dynamical
models. We illustrate the main ideas of the technique by
inferencing a model of five globally and locally coupled noisy
oscillators. Specific modifications of the technique for
inferencing hidden degrees of freedom of coupled nonlinear
oscillators is discussed in the context of physiological
applications.
\end{abstract}


\keywords{Inference, time-series analysis, cardio-respiratory
interaction, chaotic dynamics}

\section{INTRODUCTION}
 \label{sect:intro}  

Coupled oscillators are ubiquitous in nature. They are used to
describe observed phenomena intensively over the years in many
areas of science and technology including e.g. physics
\cite{Haken:83,Strogatz:94}, chemistry~\cite{Kuramoto:84} and
biology~\cite{Winfree:80}. In this approach a complex system is
characterized by projecting it onto a specific dynamical model of
coupled nonlinear oscillators. However, there are no general
methods to infer parameters of stochastic nonlinear models from
the measured time-series data. Furthermore, in a great number of
important problems the model is not usually known exactly from
\lq\lq first principles" and one is faced with a rather broad
range of possible parametric models to consider. In addition, the
experimental data can be extremely skewed, whereby important
\lq\lq hidden" features of a model such as coupling coefficients
between the oscillators can be very difficult to extract due to
the intricate interplay between noise and nonlinearity.

As was pointed out by McSharry and co-authors~\cite{McSharry:99a},
deterministic inference techniques \cite{Kantz:97} consistently
fail to yield accurate parameter estimates in the presence of
noise. The problem becomes even more complicated when both
measurement noise as well as intrinsic dynamical noise are
present~\cite{Meyer:01}. Various numerical schemes have been
proposed recently to deal with different aspects of this inverse
problem
\cite{McSharry:99a,Heald:00,Meyer:00,Friedrich:00,Meyer:01,Rossi:02,Friedrich:03}.
A standard approach to this problem is often based on optimization
of a certain cost function (a \emph{likelihood} function) at the
values of the model parameters that best reconstruct the
measurements. It can be further generalized using a Bayesian
formulation of the problem \cite{Meyer:00,Meyer:01}.
Existing techniques usually employ numerical Monte Carlo
techniques for complex  optimization \cite{Rossi:02} or
multidimensional integration \cite{Meyer:00} tasks. Inference
results from  noisy observations are shown to be very sensitive to
the  specific choice of the likelihood function
\cite{McSharry:99a}. Consequently, the \emph{correct} choice of
the \emph{likelihood function} is one of the central questions in
the inference of continuous-time noise-driven dynamical models
considered here.

In this paper, we present an efficient technique of Bayesian
inference of  nonlinear noise-driven dynamical models from
time-series data that avoids extensive numerical optimization. It
also  guarantees optimum compensation of  noise-induced errors by
invoking the likelihood function in the form of a path integral
over the random trajectories of the  dynamical system. The
technique is verified on a broad range of dynamical models
including system of five globally and locally coupled nonlinear
oscillators.

A specific example of inferencing stochastic nonlinear model from
skewed time-series data is considered in the context of
physiological research. In particular, we refer to the situation
when the variability of the cardiovascular signals is modelled in
terms of coupled nonlinear
oscillators~\cite{Saul:88,Javorka:02,Stefanovska:99a,Stefanovska:01a}.
At present there are no methods available to infer parameters of
the nonlinear coupling between oscillators directly from
experimental time series data. Furthermore, in many situations it
is important to perform such inference using univariate time
series. This rises another important issue in nonlinear
time-series analysis related to the inference of hidden dynamical
variables. If a technique of inferencing of coupling parameters
from hidden dynamical variables could be found it could provide
new effective tool for estimation of the state of autonomous
nervous control~\cite{Malpas:02} and risk stratification of
cardiovascular diseases~\cite{Leeuwen:00}. The corresponding
problem of inference of the coupling parameters of two nonlinear
oscillators perturbed by noise from univariate time-series data
will be considered in this paper.

The paper is organized as follows. In the Sec. \ref{s:theory} the
algorithm is introduced and its main features are compared with
the results of earlier work. In the Sec. \ref{s:models} the
convergence of the algorithm is analyzed in the case of inference
of coupled nonlinear stochastic oscillators. A modification of the
algorithm that allows inference of hidden dynamical variables of
two nonlinear coupled oscillators from univariate time-series data
is considered in Sec. \ref{s:univariate}.


\section{Theory of nonlinear inference of noise-driven dynamical systems}
 \label{s:theory}

Consider $N$-dimensional dynamical system described by set of
nonlinear Langevin equations
\begin{equation}
    \dot{\bf x}(t) = {\bf f}({\bf x}) + {\bf \varepsilon}(t)
    = {\bf f}({\bf x}) + {\bf\sigma}{\bf \xi}(t),
    \label{eq:dynamics}
\end{equation}
\noindent where ${\bf \varepsilon}(t)$ is an additive stationary
white, Gaussian vector noise process
\begin{equation}
    \langle {\bf \xi}(t) \rangle = 0, \quad \langle {\bf \xi}(t) \, {\bf
    \xi}^{T}(t') \rangle = {\hat {\bf D}} \, \delta(t - t'),
    \label{eq:noise}
\end{equation}
characterized by diffusion matrix ${\hat {\bf D}}$.

We assume that the trajectory $x(t)$ of this system is observed at
sequential time instants  $\lbrace t_k; k = 0, 1, \ldots, K
\rbrace$ and a series ${\cal Y} = \lbrace y_k \equiv y(t_{k})
\rbrace$ thus obtained is related to the (unknown) ``true'' system
states ${\cal X} = \lbrace x_k \equiv x( t_{k}) \rbrace$ through
some conditional PDF $p_{\rm o}\left({\cal Y}|{\cal X} \right)$.

The most general approach to dynamical model inference is based on
Bayesian framework (cf.~\cite{Meyer:01}). In the Bayesian model
inference, two distinct PDFs are ascribed to the set of unknown
model parameters: the {\em prior} $p_{\textrm{pr}}({\cal M})$ and
the {\em posterior} $p_{\textrm{post}}({\cal M}|{\cal Y})$,
respectively representing our state of knowledge about ${\cal M}$
before and after processing  a block of data ${\cal Y}$. The prior
acts as a {\em regularizer}, concentrating the parameter search to
those regions of the model space favored by our expertise and any
available auxiliary information. The two PDFs are related to each
other  via Bayes' theorem:
\begin{equation}
    \label{eq:Bayes}
    p_{\textrm{post}}({\cal M}|{\cal Y}) = \frac{{\ell}({\cal Y}|{\cal M}) \,
    p_{\textrm{pr}}({\cal M})}{\int \ell({\cal Y}|{\cal
    M}) \, p_{\textrm{pr}}({\cal M}) \, {\rm d}{\cal M}}.
\end{equation}
\noindent Here $\ell({\cal Y}|{\cal M})$, usually termed the
\emph{likelihood}, is the conditional PDF of the measurements
${\cal Y}$ for a given choice ${\cal M}$ of the dynamical model.
In practice, (\ref{eq:Bayes}) can be applied iteratively using a
sequence of data blocks ${\cal Y},{\cal Y}^{\prime}$, etc. The
posterior computed from  block ${\cal Y}$ serves as the prior for
the next block ${\cal Y}^{\prime}$, etc.  For a sufficiently large
number of observations, $p_{\textrm{post}}({\cal M}|{\cal Y},{\cal
Y}^{\prime},\ldots)$ is sharply peaked at a certain most probable
model $\cal M={\cal M}^{\ast}$.

The main efforts in the research on stochastic nonlinear dynamical
inference are focused on constructing the likelihood function that
compensates noise induced errors and on introducing efficient
optimization algorithms
(cf.~\cite{McSharry:99a,Meyer:00,Meyer:01,Rossi:02}).

No closed form expression for the likelihood function that
provides optimal compensation of the noise-induced errors was
introduced so far for continuous systems. The {\it ad hoc}
likelihood function~\cite{McSharry:99a} and their generalization
to the conditional PDF for stochastic trajectories in
maps~\cite{Meyer:00,Meyer:01} do not compensate the error in
continuous systems, since they are missing the main compensating
term (see below). The problem of noise-induced errors in inference
of continuous systems was considered in~\cite{Rossi:02} and a
general approach to constructing corresponding likelihood was
outlined. However, the closed form expression for the likelihood
that takes into account the leading compensating term was not
found and instead an {\it ad hoc} expression for the likelihood
function was used.

A common draw back of earlier research is the use of extensive
numerical optimization. This problem will become increasingly
important when complex systems with the large number (hundred or
more) of model coefficients are investigated.

In the present paper we introduce a closed form of the likelihood
function for continuous systems that provides optimal compensation
for the noise-induced errors. We also suggest parametrization of
the unknown vector field that reduces the problem of nonlinear
dynamical inference to essentially linear one for a broad class of
nonlinear dynamical systems. This allows us to write an efficient
algorithm of Bayesian inference of nonlinear noise-driven
dynamical models that avoids extensive numerical optimization and
guarantees optimum compensation of noise-induced errors.

In what follows in this section we describe the likelihood
function, the parametrization, and the corresponding algorithm.

\subsection{The likelihood function}
 \label{ss:likelihood}

It was pointed out in~\cite{Rossi:02} the probability density
functional for the nonlinear dynamical stochastic systems in
general is not known. Instead one can use the probability density
functional for random trajectories in such systems. We note that
the path-integral approach has also proved to be useful in
nonlinear filtration of random signals (see
e.g.~\cite{Rosov:02})in the situations where standard
spectral-correlation methods fail.

Therefore we write the expression for the likelihood in the form
of a path integral over the random trajectories of the system:
\begin{equation}
    \label{eq:pathint}
    \ell({\cal Y}|{\cal M}) = \int_{{\bf x}(t_{\rm i})}^{{\bf x}(t_{\rm f})} p_{\rm o}({\cal
    Y}|{\cal X}) \, {\cal F}_{\cal M}[{\bf x}(t)] \, {\cal D}{\bf x}(t),
\end{equation}
\noindent which relates the dynamical variables ${\bf x}(t)$ of
the system (\ref{eq:dynamics}) to the observations ${\bf y}(t)$.
Here we choose $t_{\rm i} \ll t_{0} < t_{K} \backslash t_{\rm f}$
so that $\ell$ does not depend on the particular initial and final
states ${\bf x}(t_{\rm i})$, ${\bf x}(t_{\rm f})$.  The form of
the probability functional ${\cal F}_{\cal M}$ over the system
trajectory ${\bf x}(t)$ is determined by the properties of the
dynamical noise ${\bf \xi}(t)$ \cite{Graham:77,Dykman:90}.

In the following we are focusing  on the case of additive and
stationary Gaussian white noise, as indicated in
(\ref{eq:dynamics}), (\ref{eq:noise}). We  consider a uniform
sampling scheme $t_{k} = t_{0} + h k$, $h \equiv (t_{K} -
t_{0})/K$ and assume  that for each trajectory component $x_n(t)$
the measurement error $\epsilon$ is negligible compared with the
fluctuations induced by the dynamical noise; that is,
$\epsilon^2\ll h ({\bf \hat D}^2)_{n\,n}$. Consequently, we  use
$p_{\rm o}({\cal Y}|{\cal X}) \rightarrow \prod_{k = 0}^{K}
\delta[{\bf y}_{k} - {\bf x}(t_k)]$ in (\ref{eq:pathint}).   Using
results from \cite{Graham:77} for ${\cal F}_{\cal M}[{\bf x}(t)]$,
the logarithm of the likelihood (\ref{eq:pathint}) takes the
following form for sufficiently large $K$ (small time step $h$):
\begin{eqnarray}
 \label{eq:likelihood}
 && \hspace{-0.3in}-\frac{2}{K}\log  \ell({\cal Y}|{\cal M})  =
 \ln\det{\hat{\bf D}}+\frac{h}{K}\sum_{k=0}^{K-1}\left[\,\,\mathop{\rm
 tr}{\hat {\bf \Phi}}({\bf y}_{k}; {\bf c}) +(\dot{\bf y}_{k}
 - {\bf f}({\bf y}_{k}; {\bf c}))^T \, {\hat {\bf D}}^{-1} \,
 \dot{\bf y}_{k}  - {\bf f}({\bf y}_{k}; {\bf c}))\right]+ N\ln(2\pi h),
\end{eqnarray}
\noindent \noindent here we introduce the \lq\lq velocity" $\dot
{\bf y}_{k}$ and matrix ${\bf \hat \Phi}({\bf x})$
\[\dot {\bf y}_{k}\equiv h^{-1} ({\bf y}_{k+1}-{\bf y}_k),\quad
({\hat {\bf \Phi}}({\bf x}; {\bf c}))_{n\,n'}\equiv
\partial f_{n}({\bf x}; {\bf c})/\partial x_{n'}.\]\noindent

It is the term $\mathop{\rm tr}{\hat {\bf \Phi}}({\bf y}_{k}; {\bf
c})$ that guarantees optimal compensation of the noise-induced
errors in our technique and that distinguish our likelihood
function from those introduced in earlier work. Formally this term
appears in path integral as a Jacobian of
transformation~\cite{Graham:73,Gozzi:83,McKane:89} from noise
variables to dynamical ones. We emphasize, however, that this term
is not a correction, but a leading term in inference as will be
shown in the following sections.

Note, that the optimization of the log-likelihood function
(\ref{eq:likelihood}) is in general essentially nonlinear problem
that requires extensive numerical optimization. Below we introduce
parametrization that allows to avoid this problem for a broad
class of nonlinear dynamical models. In particular, a vast
majority of the model examples considered in the earlier work on
the nonlinear dynamical inference can be solved using this
technique. Moreover, a large number of important practical
applications can be treated using the same approach.

\subsection{Parametrization of the unknown vector field}
 \label{ss:likelihood}

We parameterize this system in the following way.  The nonlinear
vector field ${\bf f}({\bf x})$ is written in the form
\begin{equation}
    \label{eq:model}
    {\bf f}({\bf x}) = {\hat {\bf U}}({\bf x}) \, {\bf c} \equiv {\bf f}({\bf x}; {\bf c}),
\end{equation}
\noindent where ${\hat {\bf U}}({\bf x})$ is an $N \times M$
matrix of suitably chosen basis functions $\lbrace U_{n m}({\bf
x}); \,n = 1:N,\, m = 1:M \rbrace$, and ${\bf c}$ is an
$M$-dimensional coefficient vector.

The choice of the base functions is not restricted to polynomials,
$\phi_b({\bf x})$ can be any suitable function. In general if we
use $B$ different base functions $\phi_b({\bf x})$ to model the
system (\ref{eq:dynamics}) the matrix ${\hat {\bf U}}$ will have
the following block structure
\begin{eqnarray}
 \label{eq:block_structure_1}
&&{\hat {\bf U}}=\left[
  \left[\begin{array}{llll}
    \phi_1&0&\ldots&0 \\
    0&\phi_1&\ldots&0 \\
    \vdots&\vdots&\ddots&\vdots \\
    0&0&\ldots&\phi_1 \\
  \end{array}\right]   \ldots
  \left[\begin{array}{llll}
    \phi_2&0&\ldots&0 \\
    0&\phi_2&\ldots&0 \\
    \vdots&\vdots&\ddots&\vdots \\
    0&0&\ldots&\phi_2 \\
  \end{array}\right] \ldots
  \left[\begin{array}{llll}
    \phi_B&0&\ldots&0 \\
    0&\phi_B&\ldots&0 \\
    \vdots&\vdots&\ddots&\vdots \\
    0&0&\ldots&\phi_B \\
  \end{array}\right]
  \right],
\end{eqnarray}
where we have $B$ diagonal blocks of size $N\times N$ and
$M=B\cdot N$.

An important feature of (\ref{eq:model}) for our subsequent
development is that, while possibly highly nonlinear in ${\bf x}$,
${\bf f}({\bf x}; {\bf c})$ is strictly linear in ${\bf c}$.

Eqs. (\ref{eq:likelihood}) and (\ref{eq:model}) are two main
ingredients that allow to solve problem of nonlinear stochastic
dynamical inference analytically as shown in the following
section.

\subsection{The algorithm}
 \label{ss:likelihood}

The vector elements $\lbrace c_m \rbrace$ and the matrix elements
$\lbrace D_{n n'} \rbrace$ together constitute a set ${\cal M} =
\lbrace {\bf c}, {\hat {\bf D}} \rbrace$ of unknown parameters to
be inferred from the measurements ${\cal Y}$.

With the use of (\ref{eq:model}), substitution of the prior
$p_{\textrm{pr}}({\cal M})$ and the  likelihood $\ell({\cal
Y}|{\cal M})$ into (\ref{eq:Bayes}) yields the posterior
$p_{\textrm{post}}({\cal M}|{\cal Y}) ={\rm const}\times
\exp[-S({\cal M}|{\cal Y})]$, where
\begin{equation}
    S({\cal M}|{\cal Y})\equiv S_{\textsf{y}}({\bf c},{\bf \hat D}) =
    \frac{1}{2}\rho_{\textsf{y}}({\bf \hat  D}) - {\bf c}^{T} {\bf w}_{\textsf{y}}({\bf \hat  D}) +
     \frac{1}{2}
    {\bf c}^{T} {\bf \hat  \Xi}_{\textsf{y}}({\bf \hat  D}) {\bf c}.
    \label{eq:action}
\end{equation}
\noindent Here,  use was made of the definitions
\begin{eqnarray}
  &&\hspace{-0.25in}\rho_\textsf{y}({\bf \hat{D}}) =  h \, \sum_{k = 0}^{K - 1} \dot{{\bf
   y}}_{k}^{T}   \, {\bf \hat D}^{-1} \, \dot{{\bf
   y}}_{k}  + K \, \ln (\det { \bf \hat D}),  \label{eq:defs_1} \\
 &&\hspace{-0.25in}{\bf w}_\textsf{y}({\bf \hat D}) = {\bf \hat \Sigma}_{\textrm{pr}}^{-1} \,
   {\bf c}_{\textrm{pr}} + h\sum_{k = 0}^{K - 1}\left[ {\bf \hat U}_{k}^T \,
   {\bf \hat D}^{-1} \, \dot{{\bf y}}_{k}-  \frac{{\bf v}({\bf y}_k)}{2}\right],
   \label{eq:defs_2} \\
 &&\hspace{-0.25in}{\bf \hat  \Xi}_\textsf{y}({\bf \hat  D})
    =  {\bf \hat  \Sigma}^{-1}_{\textrm{pr}} + h \, \sum_{k = 0}^{K - 1}
    {\bf \hat U}_{k}^{T} \, {\bf \hat D}^{-1} \, {\bf \hat U}_{k},
    \label{eq:defs_3}
\end{eqnarray}
\noindent where $ { \bf \hat U}_{k} \equiv {\bf \hat U}({\bf
y}_{k})$ and  the components of  vector ${\bf v}({\bf x})$ are:
\begin{equation}
\textrm{v}_{m}({\bf x})=\sum_{n=1}^{N}\frac{\partial U_{n\,m}({\bf
x})}{\partial x_n},\quad m=1:M.\label{v}
\end{equation}
\noindent

The mean values of ${\bf c}$ and ${\bf \hat D}$ in the posterior
distribution give the best estimates for the model parameters for
a given block of data ${\cal Y}$ of  length $K$ and provide a
global minimum to $ S_\textsf{y}({\bf c}, {\hat {\bf D}})$. We
handle this optimization problem in the following way. Assume for
the moment that ${\bf c}$ is known in (\ref{eq:action}). Then the
posterior distribution over ${\bf \hat D}$ has a mean  ${\bf\hat
D}^{\bf \prime}_{\textrm{post}}={\bf \hat
\Theta}_{\textsf{y}}({\bf c})$ that provides a  minimum to
$S_\textsf{y}({\bf c},{\bf \hat D})$ with respect to ${\bf \hat
D}={\bf \hat D}^T$.   Its matrix elements are
\begin{equation}
 \label{eq:updateD}
  \hspace{-0.01in}{\bf \hat \Theta}_{\textsf{y}}^{n n'}({\bf c})
  \equiv \frac{1}{K} \, \sum_{k=0}^{K-1} \left[ {\dot {\bf y}}_{k} -
    {\hat {\bf U}}({\bf y}_{k}) \, {\bf c} \right]_n \left[ {\dot {\bf y}}_{k} - {\hat {\bf U}}({\bf
    y}_{k}) \, {\bf c} \right]^{T}_{n'}.
\end{equation}
\noindent Alternatively, assume next that
 ${\hat {\bf D}}$ is known, and note
from (\ref{eq:action}) that in this case the posterior
distribution over ${\bf c}$ is Gaussian. Its covariance is given
by ${\bf \hat\Xi}_\textsf{y}({\bf \hat D})$ and the mean ${\bf
c}^{\prime}_{\textrm{post}}$ minimizes  $S_\textsf{y}({\bf c},{\bf
\hat D})$ with respect to ${\bf c}$
\begin{equation}
{\bf c}^{\prime}_{\textrm{post}}={\hat {\bf
\Xi}}^{-1}_\textsf{y}({\bf \hat D}){\bf w}_\textsf{y}({\bf \hat
D}).\label{eq:updateC}
\end{equation}
\noindent We repeat  this two-step optimization procedure
iteratively, starting from some prior values ${\bf
c}_{\textrm{pr}}$ and ${\bf \hat \Sigma}_{\textrm{pr}}$.

It can be seen that the second term in the sum on the {\it rhs} of
eq. (\ref{eq:defs_2}) originating from ${\rm tr}{\bf \hat
\Phi}({\bf y}_k)$ does not vanish at the dynamical system
attractors (\ref{eq:dynamics}), unlike the term
(\ref{eq:likelihood}) $h\sum_{k = 0}^{K - 1} {\bf \hat U}_{k}^T
\,{\bf \hat D}^{-1} \,\dot{{\bf y}}_{k}$ corresponding to the
generalized least square optimization~\cite{Theil:83}. Therefore
both types of terms are required to optimally balance the effect
of noise effect in $\{{\bf y}_k \}$ (\ref{eq:action}) and provide
the robust convergence. In the following section we analyze
relative importance of both terms for the convergence of our
algorithm.

\section{Numerical examples}
 \label{s:models}

 We verified the convergence and robustness of the algorithm on a
 broad range of dynamical systems. In this paper we will be
 specifically focused on the applications to the inference of
 coupled nonlinear oscillators.

 \subsection{Five coupled oscillators}
  \label{ss:five_coupled}

\begin{figure}[h]
   \begin{center}
   \begin{tabular}{c}
 \includegraphics[width=7cm,height=6cm]{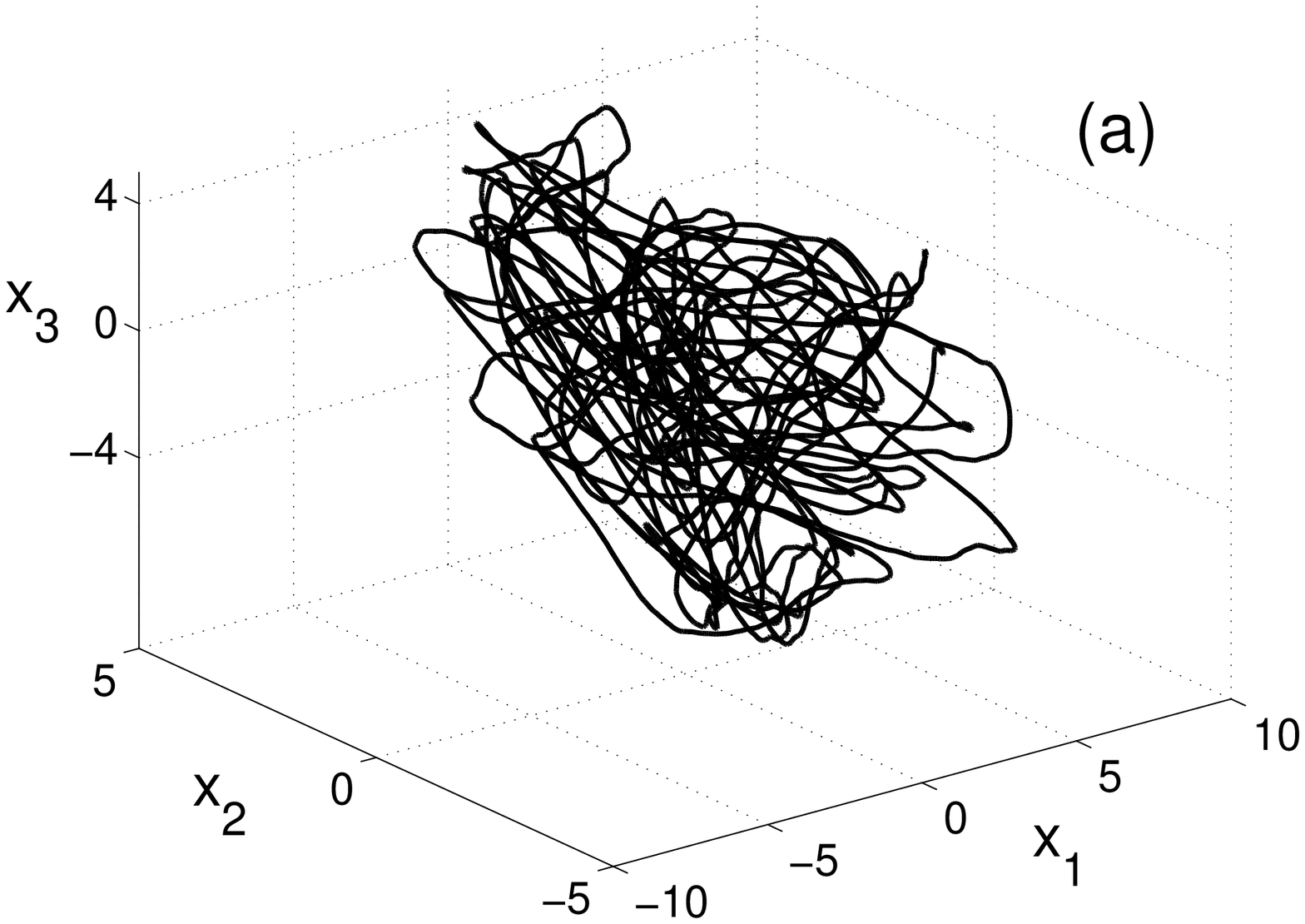}
 \hspace{1cm}\includegraphics[width=7cm,height=6cm]{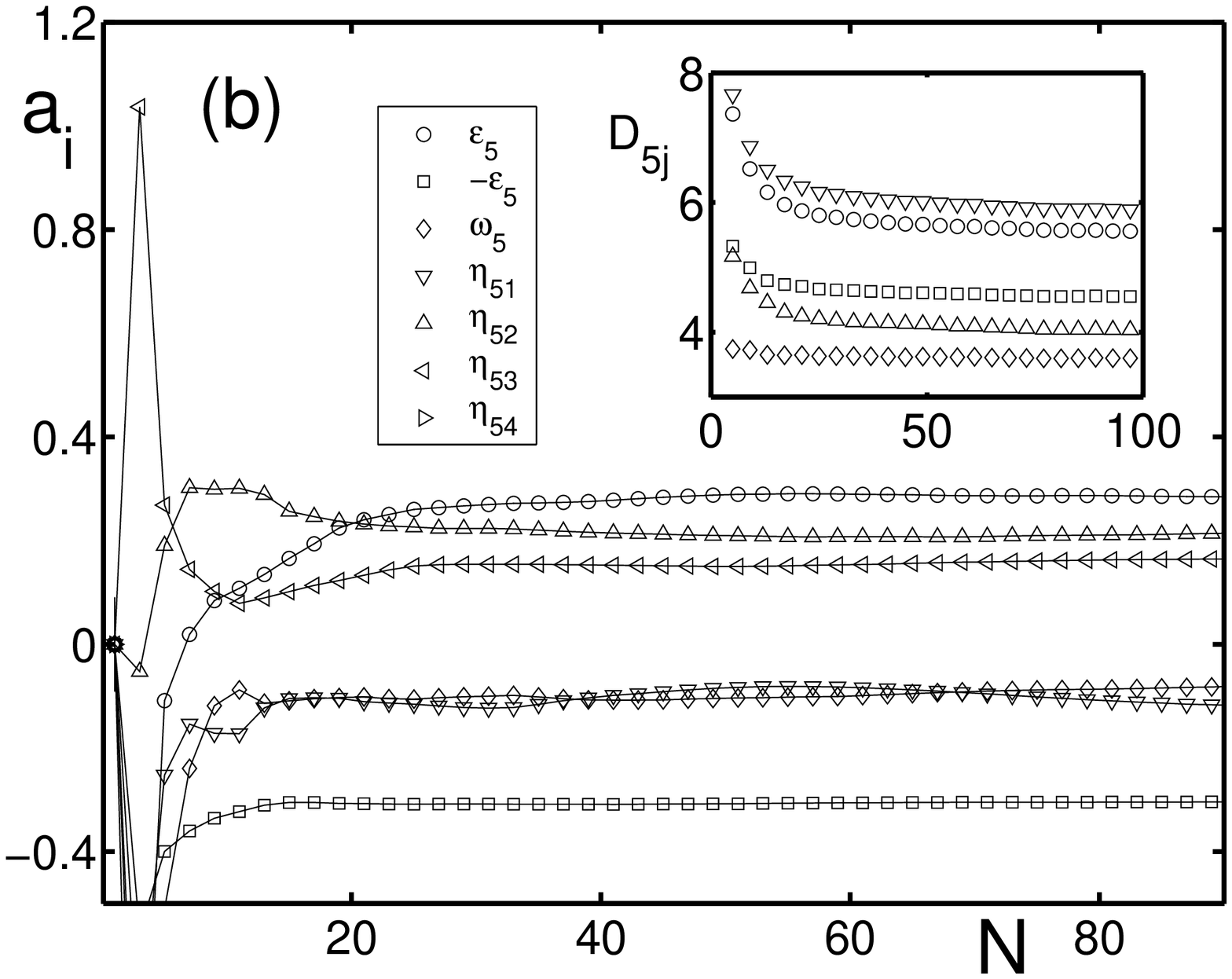}
   \end{tabular}
   \end{center}
 \caption{\label{fig:coeff_5}(a) A projection of a trajectory of system (\ref{eq:5vdp})
 on $(x_1,x_2,x_3)$ subspace of its configuration space.
 (b) Convergence of the coefficients of the 5$^{th}$ oscillator to the
 true values as a function of a number of blocks of data. We have 100 blocks of data with 800 points in each block and the
 sampling time $h=0.02$. $a_1=\epsilon_1$, $a_2=-\epsilon_1$, $a_3=-\omega_1$,
 $a_4=\eta_{12}$, $a_5=\eta_{13}$, $a_6=\eta_{14}$, $a_7=\eta_{15}$,
 $a_8=\gamma_{15}$. The convergence of the five components of the diffusion matrix is shown in the insert.}
 \end{figure}

\begin{figure}[h]
   \begin{center}
   \begin{tabular}{c}
 \includegraphics[width=9cm,height=6cm]{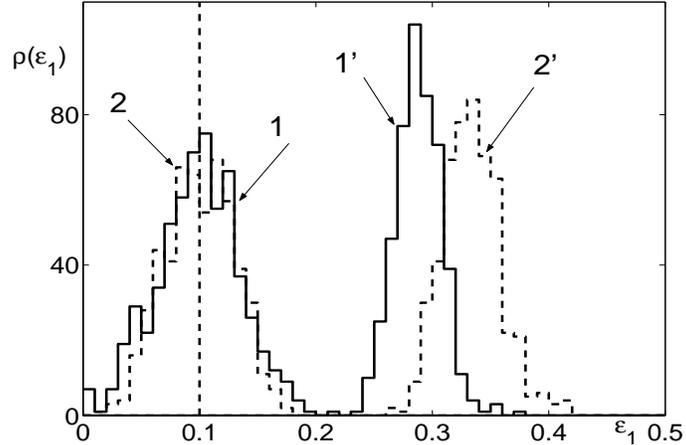}
   \end{tabular}
   \end{center}
 \caption{\label{fig:compensation} Results of inference of the
 $\varepsilon_1$ that were performed according to eqs.
 (\ref{eq:defs_2}) -- (\ref{eq:updateC}) (curves $1$ and $2$)
are compared with the results of inference without compensating
term $\mathop{\rm tr}{\hat {\bf \Phi}}({\bf y}_{k}; {\bf c})$
(curves $1'$ and $2'$) for different noise matrices $d$: $d=D$ for
(1) and (1'); $d=2*D$ for (2) and (2') where $D$ is given in eq.
(\ref{eq:D_matrix}).}
 \end{figure}

  Consider system of five locally and globally coupled van der Pol oscillators
\begin{eqnarray}
 \label{eq:5vdp}
 &&\dot x_k = y_k, \nonumber\\
 &&\dot y_k = \varepsilon_k(1-x_k^2)y_k-\omega_kx_k
 +\sum_{{j=1}\atop{j\neq k}}^5\eta_{kj}x_j
 +\gamma_{k,k-1}x_{k-1}(t)+\gamma_{k,k+1}x_{k+1}(t)+\sum_{j=1}^5\sigma_{kj}\xi_j,
\end{eqnarray}
We assume for simplicity that there is no observational noise and
that the observed signal is ${\bf y}=(y_1,y_2,y_3,y_4,y_5)$. We
note that for the model of coupled oscillators (\ref{eq:5vdp})
parameters of the equations $\dot x_k = y_k$ are known and do not
have to be inferred. An example of a trajectory of (\ref{eq:5vdp})
is shown in the figure \ref{fig:coeff_5}(a) in projection on
$(x_1,x_2,x_3)$ subspace of the configuration space of this
system.
We chose the following base functions
\[\begin{array}{ll}
      \phi(1)   = x_1;      \quad   \phi(2)   = y_1; \quad   \phi(3)   = x_2; \quad   \phi(4)   =
      y_2;\quad  \phi(5)   = x_3;      \quad   \phi(6)   = y_3;\\
      \phi(7)   = x_4;      \quad   \phi(8)   = y_4; \quad   \phi(9)   = x_5;\quad   \phi(10)   = y_5;
      \quad  \phi(11)   = x_1x_2;   \quad   \phi(12)  = x_2x_3;\\
      \phi(13)  = x_3x_4;   \quad   \phi(14)  = x_4x_5;\quad   \phi(15)  =
      x_5x_1;\quad
      \phi(16)  = x_1^2y_1;\quad   \phi(17)  = x_2^2y_2;\\
      \phi(18)  = x_3^2y_3;\quad   \phi(19)  = x_4^2y_4;\quad   \phi(20)  = x_5^2y_5.\\
   \end{array}
\]
Together with the elements of the diffusion matrix we have to
infer 115 model coefficients. Example of the convergence of the
coefficients of the 5$^{th}$ oscillator to their correct values is
shown in the Fig. \ref{fig:coeff_5}(b). Results of the
corresponding convergence for the 4$^{th}$ oscillator are
summarized in the Table \ref{tab:4th_osc}. It can be seen from the
Table that accuracy of estimation of the model parameters is
better then 1\%.
\begin{table}[h]
\caption{Convergence of the coefficients of the 4$^{th}$
oscillator of system (\ref{eq:5vdp}). We have used 200 blocks of
data with 5000 points in each block. True values are shown in the
first row, inferred values are shown in the second row, and
corresponding standard deviations are shown in the last row. The
accuracy of inference is within 5\%.} \label{tab:4th_osc}
    \begin{center}
        \begin{tabular}{|l||c|c|c|c|c|c|c|c|c|c|c|}
        \hline
\rule[-1ex]{0pt}{3.5ex} coefficients & $\varepsilon_4$ &
$\omega_4$ & $\eta_{41}$ & $\eta_{42}$ & $\eta_{43}$ &
$\eta_{44}$ & $\gamma_{43}$ & $\gamma_{45}$ & $D_{41}$ & $D_{42}$ & $D_{43}$ \\
            \hline
 \rule[-1ex]{0pt}{3.5ex}     true value & 0.2   & -0.06   & -0.075  & 0.24  & -0.23   & -0.2    & 0.064 & 0.095 & 1.477 & 2.316 & 1.783 \\   \hline
 \rule[-1ex]{0pt}{3.5ex} inferred value & 0.199 & -0.062  & -0.069  & 0.246 & -0.228  & -0.20   & 0.066 & 0.096 & 1.477 & 2.316 & 1.782 \\   \hline
\rule[-1ex]{0pt}{3.5ex}             std & 0.005 & 0.0033  & 0.0046  & 0.004 & 0.0034  & 0.0042  & 0.001 & 0.001 & 0.001 & 0.002 & 0.002 \\
            \hline
        \end{tabular}
    \end{center}
\end{table}
Of a special interest for us is the compensation of the
noise-induced errors. In the figure Fig. \ref{fig:compensation} we
compare results of inference of one of the coefficients of the
system (\ref{eq:dynamics}) $\varepsilon_1$ for two different
diffusion matrices $D$ and $2D$ where matrix $D$ was chosen at
random
\begin{eqnarray}
 \label{eq:D_matrix}
&&{\hat {\bf D}}=
  \left[\begin{array}{lllll}
    0.0621  &  1.9171 &   0.4307 &   0.0356 &   0.3113 \\
    0.5773  &  1.3597 &   0.3648 &   1.7559 &   0.3259 \\
    1.9421  &  0.1099 &   0.1535 &   0.7051 &   0.6268 \\
    1.9010  &  1.1997 &   0.0148 &   1.4443 &   0.0588 \\
    0.4561  &  0.7863 &   1.5776 &   1.9369 &   0.7153 \\
  \end{array}\right].
\end{eqnarray}
It can be shown that that without compensation term the estimator
(\ref{eq:updateC}), (\ref{eq:updateC}) is reduced to the
generalized least square (GLS) estimator. The Fig.
\ref{fig:compensation} shows that the GLS estimator systematically
overestimates the value of $\varepsilon_1$ and the larger is noise
intensity the larger is the systematic error  of the
overestimation (see curves 1' and 2' for $D$ and $2D$
correspondingly). By adding the term $\mathop{\rm tr}{\hat {\bf
\Phi}}({\bf y}_{k}; {\bf c})$ we obtain optimal compensation of
the noise-induced errors as shown by the curves 1 and 2 obtained
for the same noise intensities.
To see the effect of the compensation analytically it is
instructive to rewrite the sum in the eq. (\ref{eq:defs_2}) in the
integral form
\begin{eqnarray}
 &&\hspace{-0.25in}{\bf w}_\textsf{y}({\bf \hat D}) = {\bf \hat \Sigma}_{\textrm{pr}}^{-1} \,
   {\bf c}_{\textrm{pr}} + \int_{x(t_0)}^{x(T)}\ {\bf \hat U}({\bf y}(t))^T \,
   {\bf \hat D}^{-1} \, d{\bf y}-  \frac{1}{2}\int_{t_0}^{T}
   {\bf v}({\bf y}_k)dt,
   \label{eq:defs_2_integral}
\end{eqnarray}
It can be seen from eq. (\ref{eq:defs_2_integral}) that for the
attractor localized in the phase space the first integral is
finite, since initial and final points of integration belong to
the attractor. The second integral is growing when the total time
of inference is growing.

In particular, for a point attractor this integral is identically
zero and the whole inference is due to the compensating term
$\frac{1}{2}\int_{t_0}^{T}{\bf v}({\bf y}_k)dt$. This result is
intuitively clear, since for the point attractor in the absence of
noise the system will stay forever in the same point and no
inference can be done. It is only noise that forces the system to
move about in the phase space and makes it possible to perform
inference.

\section{Inference of two coupled oscillators from univariate time-series data}
 \label{s:univariate}

As we have mentioned in the introduction in many real experimental
situations the model is not usually known exactly from \lq\lq
first principles" and in addition, the experimental data can be
extremely skewed, whereby important \lq\lq hidden" features of a
model such as coupling coefficients can be very difficult to
extract due to the intricate interplay between noise and
nonlinearity.

A specific example of such experimental situation is inference of
the strength, directionality and a degree of randomness of the
cardiorespiratory interaction from the blood pressure signal. Such
inference can provide valuable diagnostic information about the
responses of the autonomous nervous system
\cite{Hayano:03,Malpas:02}. However, it is inherently difficult to
dissociate a specific response from the rest of the cardiovascular
interactions and the mechanical properties of the cardiovascular
system in the intact organism~\cite{Jordan:95}. Therefore a number
of numerical techniques were introduced to address this problem
using e.g. linear approximations~\cite{Taylor:01}, or
semi-quantitative estimations of either the strength of some of
the nonlinear terms~\cite{Jamsek:03} or the directionality of
coupling~\cite{Rosenblum:02,Palus:01}. But the problem remains
wide open because of the complexity and nonlinearity of the
cardiovascular interactions.

It is important to notice that simultaneous measurements of the
cardiovascular signals is performed in different parts of the
system (see e.g.~\cite{Stefanovska:99a}). As a consequence the
nonlinear characteristics of the oscillations are substantially
modified in each signal and inference of nonlinear coupling
parameters has to be performed preferably using univariate data
e.g. blood pressure or blood flow signal only. The necessity to
use univariate data in general poses serious limitations on the
techniques of reconstruction and the problem become essentially
nontrivial even in quasi-linear noise-free
limit~\cite{Janson:01,Janson:02a,Janson:02b}.

\begin{figure}[h]
   \begin{center}
   \begin{tabular}{c}
 \includegraphics[width=12cm,height=4cm]{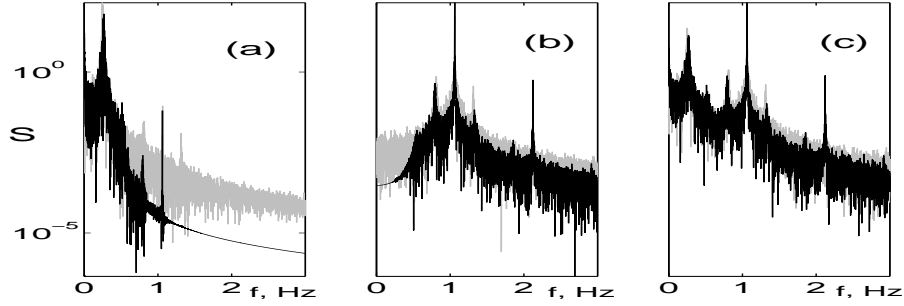}
   \end{tabular}
   \end{center}
 \caption{\label{fig:univariate_spectra} Comparison of the power spectra of the
 inferred $z(t)$ components of the signal (gray lines) with the original signal $s(t)$ (black lines):
 (a) a low-frequency component of the signal $s_0(t)$ obtained using low-pass Butterworth filter of the
 5$^{th}$ order with cut-off frequency 0.55 Hz (black line) is compared with the inferred signal $z_0(t) (gray line)$;
 (b) a high-frequency component of the signal $s_1(t)$ obtained high-pass Butterworth filter of the 4$^{th}$
 order with cut-off frequency 0.55 Hz (black line) is compared with the inferred signal $z_1(t)$ (gray line);
 (c) spectrum of the original signal $s(t)=x_1(t)+x_2(t)$ (black line) is compared with the spectrum of the
 inferred signal $z(t)=z_0(t)+z_1(t)$.}
 \end{figure}

In this section we investigate the possibility of extending our
technique of reconstruction of coupled nonlinear stochastic
oscillators to encompass the case of inference from the univariate
time-series data in the context of physiological research. We note
that this is a particular example of inference of hidden dynamical
variables, which will be addressed elsewhere.
 \begin{figure}[h]
   \begin{center}
   \begin{tabular}{c}
 \includegraphics[width=12cm,height=4cm]{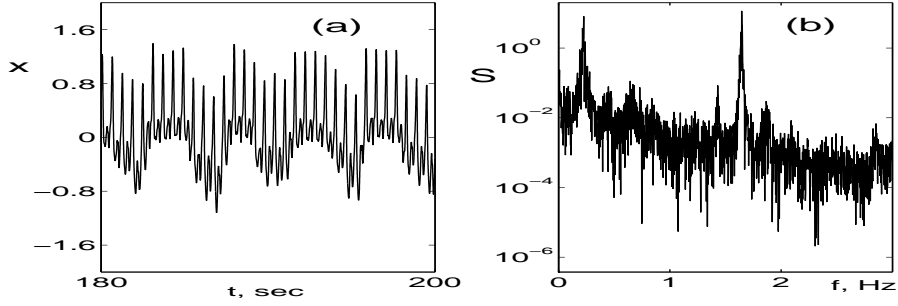}
   \end{tabular}
   \end{center}
 \caption{\label{fig:real_data} Example of the blood pressure
 signal (a) and of its spectrum (b) taken from the record 24 of the
 MGH/MF Waveform Database available at www.physionet.org.}
 \end{figure}
An example of the actual signal of the central venous blood
pressure (record 24 of the MGH/MF Waveform Database available at
www.physionet.org). The main features of the blood signal data is
the presence of the two oscillatory components at frequencies
approximately $f_r=0.2$ Hz and $f_c=1.7$ Hz corresponding to the
respiratory and cardiac oscillations. It can also be clearly seen
from the spectra that the nonlinear terms including terms of
nonlinear interaction (and cardiorespiratory interaction in
particular) are very strong in this sample. We note that the
relative intensity and position of the cardiac and respiratory
components vary strongly from sample to sample with average
frequency of the respiration being around $0.3$ Hz and of the
heart beat being around $1.1$ Hz. To infer coupling parameters
from the univariate blood pressure signal an important simplifying
assumption can be used. Namely it is assumed that the blood
pressure signal can be represented as the sum of the oscillatory
components with the main contributions coming from the
oscillations of the respiration and heart~\cite{Stefanovska:99a}.
Accordingly we chose our surrogate data as a sum of coordinates of
two coupled van der Pol oscillators $s(t) = x_1(t)+x_2(t)$. It can
be seen that the spectrum of $s(t)$ (Fig.
\ref{fig:univariate_spectra} (c)) reproduces mentioned above main
features of the real blood pressure signal.
\begin{eqnarray}
\label{eq:2vdp}
 &&\dot x_1 = y_1, \qquad \dot y_1 = \epsilon_1(1-x_1^2)y_1-\omega_1^2x_1
 + \alpha_1x_2 + \sum_{i,j=1}^{2}\beta_{1,ij}x_ix_j
 + \sum_{{i,j=1}\atop{j\neq i}}^{2}\gamma_{1,ij}x_iy_j
 + \sum_{j=1}^{2}\sigma_{1j}\xi_j, \\
 &&\dot x_2 = y_2,  \qquad \dot y_2 =
 \epsilon_2(1-x_2^2)y_2-\omega_2^2x_2
 + \alpha_2x_2 + \sum_{i,j=1}^{2}\beta_{2,ij}x_ix_j
 + \sum_{{i,j=1}\atop{j\neq i}}^{2}\gamma_{2,ij}x_iy_j
 + \sum_{j=1}^{2}\sigma_{2j}\xi_j,\\
 && \langle\xi_i(t)\rangle = 0, \qquad \langle\xi_i(t)\xi_j(t')\rangle =
 \delta_{ij}\delta(t-t').\nonumber
\end{eqnarray}
Here noise matrix $\sigma$ mixes zero-mean white Gaussian noises
$\xi_j(t)$ and is related to the diffusion matrix
$D=\sigma\cdot\sigma^T$.

\begin{figure}[h]
   \begin{center}
   \begin{tabular}{c}
 \includegraphics[width=12cm,height=4.5cm]{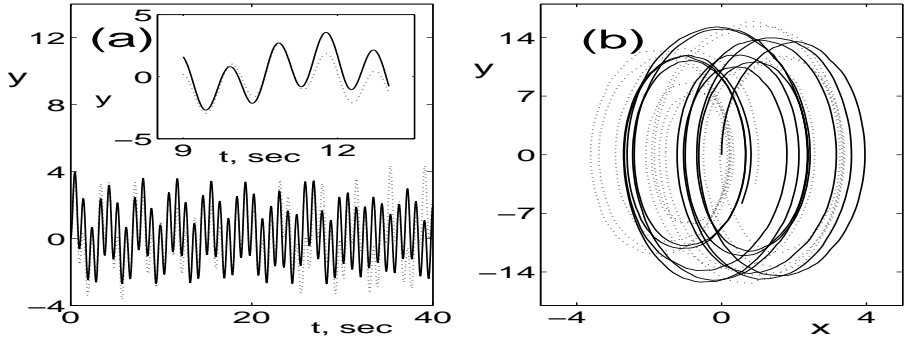}
   \end{tabular}
   \end{center}
 \caption{\label{fig:univariate_xy} (a) Comparison of the inferred signal
 $x(t)=x_1(t)+x(2)$ (black solid line) with the original signal (black
 dotted line). In the insert fragments of both signals are compared with
 better resolution in time. (b) Comparison of the inferred phase space
 trajectory $(x(t), y(t))$ (black solid line) with the original one
 (black dotted line). To facilitate the comparison we have used the same
 initial conditions to generate phase space trajectory with exact parameters
 of the system and with inferred parameters of the system.}
 \end{figure}

To infer parameters of nonlinear coupling between cardiac and
respiratory oscillations we decompose \lq\lq measured'' signal
$s(t)$ on two oscillatory components using a combination of low-
and high-pass Butterworth filters representing observations of
mechanical cardiac and respiratory degrees of freedom on a
discrete time grid with step $h=0.02$ sec. Obtained in this way
signals $s_0(t)$ and $s_1(t)$ are shown in the Fig.
\ref{fig:univariate_spectra} (a) and (b)
respectively.\footnote{Note the difference between actual
oscillatory components of the original signal $s(t)$: $x_1(t)$ and
$x_2(t)$ and the components obtained using spectral decomposition
with Butterworth filters: $s_0(t)$ and $s_1(t)$.} To make this
numerical experiment even more realistic the input signal $s(t)$
was filtered before decomposition (using high-pass Butterworth
filter of the 2$^{nd}$ order with cut-off frequency 0.0025 Hz) to
model standard procedure of de-trending, which is used in
time-series analysis of the cardiovascular signals to remove low
frequency non-stationary trends.
We now use standard embedding procedure to introduce an auxiliary
two-dimensional dynamical system whose trajectory ${\bf
z}(t)=(z_0(t),z_1(t))$ is related to the observations $\{{\bf
s}(t_k)\}$  as follows
\[z_n(t_k) = \frac{s_n(t_k+h)-s_n(t_k)}{h},\]
where $n=1,2$. The corresponding simplified model of the nonlinear
interaction between the cardiac and respiratory limit cycles has
the form (cf. with ~\cite{Stefanovska:01a})
\begin{eqnarray}
 \label{eq:embedded}
 &&\dot z_n =
 b_{1,n}s_n+b_{2,n}s_{n-1}+b_{3,n}z_n+b_{4,n}z_{n-1}+b_{5,n}s_n^2+b_{6,n}s_{n-1}^2
 + b_{7,n}z_n^2+b_{8,n}z_{n-1}^2 \nonumber \\
 &&+ \,b_{9,n}s_ns_{n-1}+b_{10,n}s_nz_n+b_{11,n}s_nz_{n-1}+b_{12,n}s_{n-1}z_n
 +b_{13,n}s_{n-1}z_{n-1}+ b_{14,n}z_nz_{n-1} \nonumber\\
 &&+\, b_{15,n} s_n^2 z_n^2+ b_{16,n} s_{n-1} z_n^2 +\xi_{n}(t),
 \quad n=0,1.
\end{eqnarray}
\noindent where $\xi_n(t)$ is a  Gaussian white noise with
correlation matrix $Q_{n\,n'}$. We emphasize that a number of
important parameters of the decomposition of the original signal
(including the bandwidth, the order of the filters) have to be
selected to minimize the cost (\ref{eq:action}) and provide the
best fit to the measured time series $\{ {\bf s}(t_k)\}$. The
parameters of the model (\ref{eq:embedded}) can now be inferred
directly from the univariate \lq\lq measured'' time series data
$s(t)$. The comparison between the time series of the inferred and
actual cardiac oscillations is shown in  Fig.
\ref{fig:univariate_spectra} and Fig. \ref{fig:univariate_xy}.
The comparison of the inferred parameters with their actual values
is summarized in the Tables \ref{tab:univariate}.
\begin{table}[!h]
\caption{Comparison of the non-zero parameters of the second
equation of the model (\ref{eq:embedded}) inferred from the
univariate time series data $s(t)$ with their actual values. We
have used 1 block of data with 120000 points. True values are
shown in the first row, inferred values are shown in the second
row.} \label{tab:univariate}
    \begin{center}
        \begin{tabular}{|l||c|c|c|c|c|c|c|c|c|c|}
        \hline
 \rule[-1ex]{0pt}{3.5ex} coefficients
   & $b_{2,1}$ & $b_{2,2}$ & $b_{2,3}$ & $b_{2,4}$ & $b_{2,5}$ & $b_{2,6}$ & $b_{2,9}$ & $b_{2,14} $& $b_{2,16}$ & $D_1$ \\
            \hline
 \rule[-1ex]{0pt}{3.5ex} true value & 0.05 & -45.0 & -0.19 & 0.25 & 0 & 2.55 & 0.2 & 0.11 & -0.25 & 0.2\\
            \hline
 \rule[-1ex]{0pt}{3.5ex} inferred value & 0.014  & -44.73 & -0.071 & 0.17 & -0.081 & 1.25 & 0.415 & 0.14 & -0.251 & 0.17\\
            \hline
        \end{tabular}
    \end{center}
\end{table}
It can be seen from the Table that the inferred parameters give
correct order of the magnitude for the actual values. The inferred
values can be further corrected taken into account attenuation of
the filters at different frequencies. We emphasize, however, that
the technique of spectral decomposition of the \lq\lq measured''
signal is in principal non-unique. Moreover, in the actual
experimental situation the dynamics of the physiological
oscillations is unknown and can be only very approximately
modelled by the system of coupled oscillators. Furthermore, the
only criterion for the goodness of the spectral decomposition is
the coincidence of the original and inferred signal and spectrum.
For these reasons the estimation of the model parameters with the
accuracy better then the order of magnitude does not improve the
quality of the inferred information as will be discussed in more
details elsewhere.

In conclusion, we suggested new technique of inference of
parameters nonlinear stochastic dynamical system. The technique
does not require extensive global optimization, provides optimal
compensation for noise-induced errors and is robust in a broad
range of dynamical models. We illustrate the main ideas of the
technique by inferencing 115 model coefficients of five globally
and locally coupled noisy oscillators within accuracy 1\%. It is
demonstrated that our technique can be readily extended to solve
selected problems of nonlinear stochastic inference of hidden
dynamical variables in the context of the physiological modelling.
We show in particular that the method allows one to estimate
correct order of the magnitude of nonlinear coupling of two
stochastic oscillators from univariate time series data. The
framework of nonlinear Bayesian inference outlined in this paper
can be further generalized to include errors of measurements and
to solve problem of global inference hidden dynamical variables.

\subsection{Acknowledgments}
The work was supported by the Engineering and Physical Sciences
Research Council (UK), NASA  IS IDU project (USA),  the Russian
Foundation for Fundamental Science, and INTAS.



\end{document}